\documentclass[showpacs,amsmath,amssymb,twocolumn,pra,superscriptaddress,notitlepage]{revtex4-1}

\usepackage{qcircuit}
\usepackage{amsmath,bm}
\usepackage[dvips]{graphicx}
\usepackage{amsmath,amssymb,amsthm,mathrsfs,amsfonts,dsfont}
\usepackage{braket}
\usepackage{bm}
\usepackage{enumerate}
\usepackage{color}
\usepackage{graphicx}
\usepackage{appendix}
\usepackage{algorithm}
\usepackage{algorithmic}
\usepackage[justification=RaggedRight]{caption}
\usepackage[subrefformat=parens]{subcaption}
\begin{document}


\title{\textcolor{black}{Variational secure cloud  quantum computing} 
}

\author{Yuta Shingu}
\email{shingu.yuta@aist.go.jp}
\affiliation{Department  of  Physics,  Faculty  of  Science  Division  I,Tokyo  University  of  Science,  Shinjuku,  Tokyo  162-8601,  Japan.}
\affiliation{Device Technology Research Institute,  National  Institute  of  Advanced  Industrial  Science  and  Technology  (AIST),1-1-1  Umezono,  Tsukuba,  Ibaraki  305-8568,  Japan.}

\author{Yuki Takeuchi}
\email{yuki.takeuchi.yt@hco.ntt.co.jp:Y. S. and Y. T.  contributed to this
work equally} 
\affiliation{NTT Communication Science Laboratories, NTT Corporation,3-1 Morinosato Wakamiya, Atsugi, Kanagawa 243-0198, Japan}

\author{Suguru Endo}
\affiliation{NTT Secure Platform Laboratories, NTT Corporation, Musashino 180-8585, Japan.}

 \author{Shiro Kawabata}
 \affiliation{Device Technology Research Institute,  National  Institute  of  Advanced  Industrial  Science  and  Technology  (AIST),1-1-1  Umezono,  Tsukuba,  Ibaraki  305-8568,  Japan.}

 \author{Shohei Watabe}
 \affiliation{Department  of  Physics,  Faculty  of  Science  Division  I,Tokyo  University  of  Science,  Shinjuku,  Tokyo  162-8601,  Japan.}

 \author{Tetsuro Nikuni}
  \email{nikuni@rs.kagu.tus.ac.jp}
 \affiliation{Department  of  Physics,  Faculty  of  Science  Division  I,Tokyo  University  of  Science,  Shinjuku,  Tokyo  162-8601,  Japan.}

\author{Hideaki Hakoshima}
\affiliation{Device Technology Research Institute,  National  Institute  of  Advanced  Industrial  Science  and  Technology  (AIST),1-1-1  Umezono,  Tsukuba,  Ibaraki  305-8568,  Japan.}

\author{Yuichiro Matsuzaki
}
\email{matsuzaki.yuichiro@aist.go.jp}  
\affiliation{Device Technology Research Institute,  National  Institute  of  Advanced  Industrial  Science  and  Technology  (AIST),1-1-1  Umezono,  Tsukuba,  Ibaraki  305-8568,  Japan.}


\begin{abstract}
Variational quantum algorithms (VQAs) have been considered to be useful applications of noisy intermediate-scale quantum (NISQ) devices. Typically, in the VQAs, a parametrized ansatz circuit is used to generate a trial wave function, 
and the parameters are optimized to minimize a cost function. 
On the other hand, blind quantum computing (BQC) has been studied \textcolor{black}{in order to 
provide the quantum algorithm} with security \textcolor{black}{by using} cloud networks. A client with a limited ability to perform quantum operations hopes to have access to a quantum computer of a server, and BQC allows the client to use the server's computer without leakage of the client's information (such as \textcolor{black}{input, running quantum algorithms}, and output) to the server. However, BQC is designed for fault-tolerant quantum computing, and this requires many ancillary qubits, which may not be suitable for NISQ devices. Here, we propose an efficient way to implement the NISQ computing with guaranteed security for the client. In our architecture, only $N+1$ qubits are required, \textcolor{black}{under an assumption that the form of ansatzes is known to the server}, where $N$ denotes the necessary number of the qubits in the original NISQ algorithms.
The client only performs single-qubit measurements on an ancillary qubit sent from the server, and the measurement angles can specify the parameters for the ansatzes of the NISQ algorithms.
No-signaling principle guarantees that neither parameters chosen by the client nor the outputs of the algorithm are leaked to the server. This work paves the way for new applications of NISQ devices.

\end{abstract}

\maketitle

\section{Introduction}

Quantum devices have the potential to offer significant advantages over classical devices. 
Entanglement and superposition play an essential role in the quantum advantage.
Especially, quantum computation, quantum cryptography, and quantum metrology are considered promising applications of quantum devices~\cite{doi:10.1137/S0097539795293172,grover1997quantum,harrow2009quantum,vandersypen2001experimental,bennett1984proceedings,bennett1992experimental,RevModPhys.74.145,5438603,dowling2003quantum,doi:10.1080/00107510500293261,RevModPhys.89.035002,budker2007optical,balasubramanian2008nanoscale,maze2008nanoscale,neumann2013high,wineland1992spin,huelga1997improvement,matsuzaki2011magnetic,chin2012quantum}.
Quantum computation provides faster calculations than the classical one~\cite{doi:10.1137/S0097539795293172,grover1997quantum,harrow2009quantum,vandersypen2001experimental}. 
Quantum cryptography ensures information-theoretic security for the communication between distant sites~\cite{bennett1984proceedings,bennett1992experimental,RevModPhys.74.145}. Quantum metrology aims to \textcolor{black}{create} a superior sensor to a classical one by using entanglement~\cite{RevModPhys.89.035002,budker2007optical,balasubramanian2008nanoscale,maze2008nanoscale,neumann2013high,wineland1992spin,huelga1997improvement,matsuzaki2011magnetic,chin2012quantum}.

Recently, great efforts have been devoted
to  the hybridization between quantum computation, quantum cryptography, and quantum metrology~\cite{kessler2014quantum,dur2014improved,arrad2014increasing,herrera2015quantum,unden2016quantum,matsuzaki2017magnetic,higgins2007entanglement,waldherr2012high,komar2014quantum,proctor2018multiparameter,eldredge2018optimal}. A technique of quantum 
computation such as error correction
or phase estimation algorithm has been used in quantum sensing to improve sensitivity~\cite{kessler2014quantum,dur2014improved,arrad2014increasing,herrera2015quantum,unden2016quantum,matsuzaki2017magnetic} and/or dynamic range~\cite{higgins2007entanglement,waldherr2012high}. Quantum network can be combined with quantum sensing to detect global information of the target fields~\cite{komar2014quantum,proctor2018multiparameter,eldredge2018optimal}\textcolor{black}{, and to add security about the sensing target~\cite{giovannetti2002quantum,giovannetti2002positioning,chiribella2007secret,huang2019cryptographic,takeuchi2019quantum,yin2020experimental}}.

Especially, blind quantum computation (BQC) is an 
\textcolor{black}{idea to combine}
quantum computation and quantum cryptography~\cite{broadbent2009universal,PhysRevA.87.050301,PhysRevA.93.052307,Barz303,Greganti_2016}. Suppose that a client who does not have a sophisticated quantum device hopes to access a server that has \textcolor{black}{a scalable fault-tolerant} quantum computer.
The BQC provides a client with a way to access the server's quantum computer in a secure way where the client's information such as input, output, and algorithm is not leaked to the server.
The key idea of the BQC is to use measurement-based quantum computation (MBQC)\cite{raussendorf2001one,raussendorf2003measurement,walther2005experimental}. In the MBQC, a cluster state is generated as a resource of the entanglement, and then a sequence of single-qubit measurements is performed. Depending on the algorithm, one needs to change angles of the single-qubit measurements, while the form of the cluster state does not depend on the choice of the algorithm. If the server sends a cluster state to the client and the client performs the single-qubit measurements, the server does not obtain any information of either the details or output of the algorithm set by the client. The no-signaling principle guarantees the security of the protocol~\cite{popescu1994quantum,morimae2013blind}.

Recently many theoretical and experimental works have been devoted to developing quantum devices in the noisy intermediate-scale quantum (NISQ) era. The NISQ device could involve tens to thousands of qubits with a gate error rate of around $10^{-3}$ \cite{endo2021hybrid}.
The NISQ computing typically requires
only a shallow circuit to implement quantum algorithms. 
Variational quantum algorithms (VQAs) are the typical application of the NISQ computing~\cite{peruzzo2014variational,kandala2017hardware,moll2018quantum,mcclean2016theory,farhi2014quantum,li2017efficient,yuan2019theory}.
In the VQA, one generates a trial wave function from a parametrized ansatz circuit that is typically shallow. In order to optimize a cost function tailored to a problem, one updates the parameters with classical computation to generate a new 
trial wave function.
One can search exponentially large Hilbert space with the parametzied quantum circuit via the repetition of such hybrid quantum-classical operations, and thus \textcolor{black}{could} 
find a solution to a given problem. 


A natural question is whether one can implement the NISQ computing in the blind architecture. If one adopts the BQC with the MBQC, one can in principle perform any gate-type quantum computation including NISQ computing. However, to implement the BQC with the MBQC on the cluster state, the necessary number of the qubits is around $3N$~\cite{raussendorf2001one,raussendorf2003measurement,walther2005experimental}, where $N$ is the number of the qubits required in the original NISQ algorithm. Since the number of the qubits in the blind architecture with the MBQC is much larger than that in the original algorithm without \textcolor{black}{blind properties}~\cite{broadbent2009universal,PhysRevA.87.050301,PhysRevA.93.052307,Barz303,Greganti_2016}, such a scheme may not be implementable with the NISQ device with a limited number of qubits.

Here, we propose an efficient scheme to implement the 
\textcolor{black}{variational secure cloud  quantum computing.}
The purpose of our scheme is that the client accesses the quantum computer of the server to implement the NISQ computing in a secure way where the information of the ansatz circuit's parameters 
and output of the algorithm are not leaked to the server. 
\textcolor{black}{This is essential for  security, because the ansatz circuit's parameters could contain important information such as private data especially when we perform machine learning with NISQ devices \cite{mitarai2018quantum,schuld2020circuit,farhi2020classification,zoufal2021variational,shingu2020boltzmann}.}
Importantly, our scheme requires only $N+1$ qubits while MBQC on the cluster state requires around $3N$ qubits.
The key idea of our scheme is to use an ancillary qubit for the implementation of the quantum gates on register qubits of the server.
The server performs
only a limited set of gate operations with fixed angles, \textcolor{black}{ namely,} \textcolor{black}{Hadamard} operations and controlled-$Z$ gates
on the register qubits,
while the client performs arbitrary single-qubit measurements on the
ancillary qubit.

\textcolor{black}{A key idea of our scheme is the use of ancilla-driven quantum computation (ADQC)~\cite{anders2010ancilla,bocharov2015efficient,browne2016one,paetznick2013repeat}.
The ADQC was originally discussed as one of the novel ways to perform the gate-type quantum computation.
\textcolor{black}{We discuss, for the first time to our best knowledge, } the use of the ancilla-driven architecture for 
NISQ computing with security inbuilt.}
In our architecture, the server couples an ancillary qubit to a register qubit
via a fixed two-qubit gate at the server side,  and the
ancillary qubit is sent to the client. Then the client implements single-qubit measurements on the
ancillary qubit.
By repeating this process, the client can specify the parameters for the NISQ computing by the angles of the
single-qubit measurements, and  also can obtain the output of the algorithm from the readout of the ancillary qubit. Importantly, in this scheme, the client does not send any qubits nor classical signals to the server, and thus both client's operations and measurement results are unknown to the server.
Therefore, the information about the parameters and output of the NISQ algorithm cannot be leaked to the server due to the no-signaling principle \cite{popescu1994quantum,morimae2013blind}. 

The paper is structured as follows. In Secs. II and III, we review the ADQC and NISQ algorithm, respectively. In Sec. IV, we describe our architecture of the NISQ computing with security inbuilt.
In Sec. V, we conclude our results.

\section{Ancilla-driven quantum computation}
In the ADQC~\cite{anders2010ancilla}, 
we define register qubits to execute algorithms, and also define an ancillary qubit that can be spatially transferred from one place to another. The basic idea of the ADQC is to entangle the register qubit and ancillary qubit, and the ancillary qubit is sent to another place for the measurement at a specific angle. These operations allow one to perform a universal set of operations. For the implementation with the physical systems,
register qubits can be solid-state systems
that can interact with photons, and the ancillary qubit can be an optical photon that is transmitted to a distant place.

\subsection{Single-qubit rotation on a register qubit}
\label{subsection:Single qubit rotation on a register qubit}
\begin{figure}
  \includegraphics[width=0.45\textwidth]{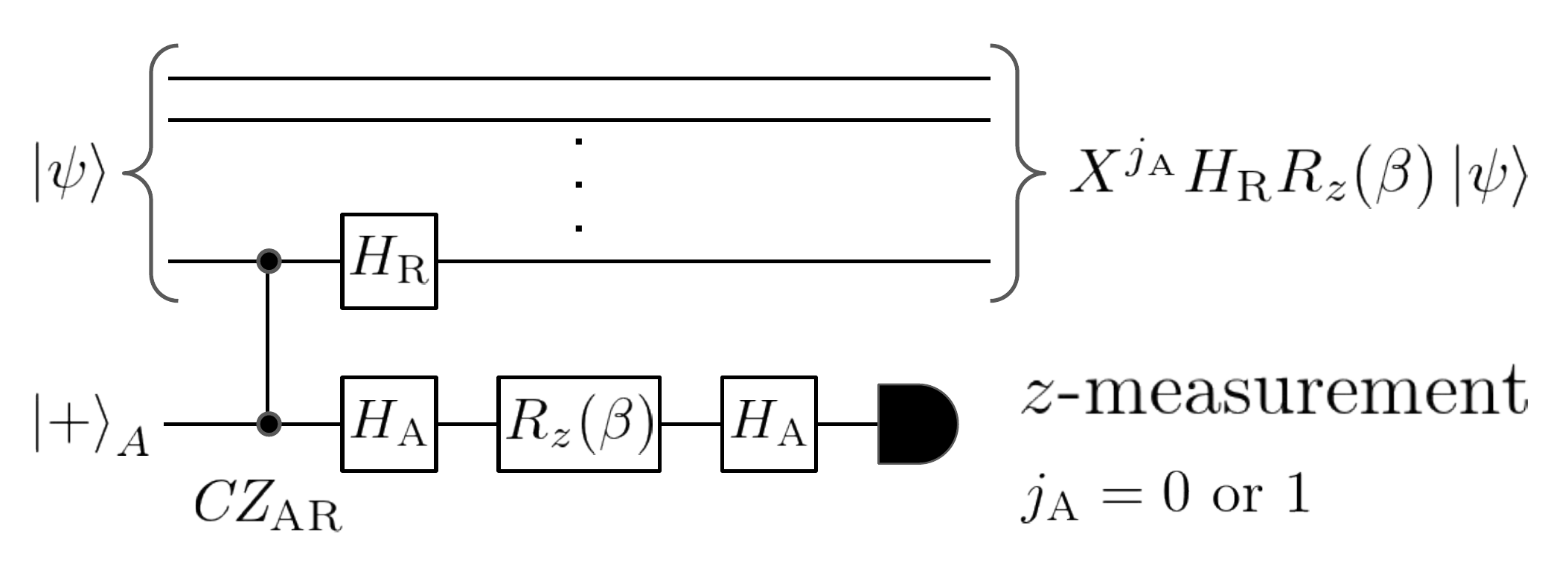}
  \caption{
  The circuit for implementing the ancilla-driven quantum computation (ADQC). The upper horizontal lines (the lower line) represents register qubits (an ancillary qubit), where $\mbox{\it CZ}_{\mathrm{AR}}$ denotes the controlled-$Z$ gate between \textcolor{black}{one of the register qubits} and the ancillary qubit, $H_{\mathrm{R}}$ ($H_{\mathrm{A}}$) denotes the Hadamard gate for the register (ancillary) qubit, and $R_z(\beta)$ denotes a parameterized rotation around the $z$-axis with any value $\beta$.
  We prepare the initial states $\ket{\psi}$ and $\ket{+}_{\mathrm{A}}$ for the register qubits and the ancillary qubit, respectively, where $\ket{\psi}$ denotes an 
  arbitrary input state. 
  After implementing the unitary operation $H_{\mathrm{A}} R_z(\beta) E_{\mathrm{AR}}$ and measuring the ancillary qubit in the $z$-basis, \textcolor{black}{where $E_{\mathrm{AR}}\equiv H_{\mathrm{A}} H_{\mathrm{R}} \mbox{\it CZ}_{\mathrm{AR}}$,}  we obtain $X^{j_{\mathrm{A}}} H_{\mathrm{R}} R_z(\beta)\ket{\psi}$ at the register qubits, where $X$ denotes the Pauli-$X$ gate and $j_{\mathrm{A}}=0 \ \rm{or} \ 1$ is the result of the measurement. }
  \label{fig:ADQC}
\end{figure}

We explain a realization of single-qubit rotation along $z$-axis \textcolor{black}{as follows} (see Fig.~\ref{fig:ADQC}).
\textcolor{black}{
\begin{enumerate}
\item We prepare a state $\ket{+}_{\mathrm{A}}\equiv\frac{1}{\sqrt{2}}(\ket{0}+\ket{1})$ of an ancilla qubit (which we call qubit A)  
and any state $\ket{\psi}$ of register qubits (which we call qubits R).
\item The ancillary qubit A is coupled with one of the register qubits R
via a controlled-$Z$ gate $\mbox{\it CZ}_{\mathrm{AR}}$, and subsequently, we implement two Hadamard gates $H_{\mathrm{A}}$ and $H_{\mathrm{R}}$ to the qubit A and the qubit R, respectively. Thus, we have a unitary operation of $E_{\mathrm{AR}}\equiv H_{\mathrm{A}} H_{\mathrm{R}} \mbox{\it CZ}_{\mathrm{AR}}$.
\item A rotation about the $z$-axis $R_z(\beta)$ and a Hadamard gate are implemented on the ancillary qubit, where $\beta$ is an arbitrary rotation angle.
\item Measuring the ancillary qubit in the $z$-basis projects the state of the register qubit onto \textcolor{black}{$X^{j_{\rm{A}}} H_{\mathrm{R}} R_z(\beta)\ket{\psi}$, where $j_{\rm{A}}=0 \ {\rm or} \ 1$} is the result of the measurement \textcolor{black}{on the ancillary qubit}.
\end{enumerate}
}
The third and the last steps can be unified into a single measurement step if an arbitrary-angle single-qubit measurement can be implemented on 
\textcolor{black}{the ancillary qubit.}
\textcolor{black}{The details of performing an arbitrary single-qubit rotation and two-qubit gates with ADQC are explained in Appendix~\ref{detailed ADQC}.}

\section{Variational quantum algorithm for NISQ device}

\textcolor{black}{Variational quantum algorithm (VQAs)} perform a required task by preparing a parametrized wave function on a quantum circuit $\ket{\psi(\vec{\theta})}$ with the variational parameters $\vec{\theta}$ to be optimized by minimizing a cost function $C(\vec{\theta})$ tailored to a problem. The parametrized wavefunction can be generally described as $\ket{\psi(\vec{\theta})}=U_{\rm{AN}}(\vec{\theta}) \ket{\bar{0}}$ \textcolor{black}{with  $\ket{\bar{0}} \equiv \bigotimes_{i=1}^{N}\ket{0}$}, where the ansatz quantum circuit is represented as a repetition of parametrized quantum gates and fixed quantum gates as $U_{\rm{AN}}(\vec{\theta})= V_{L+1}U_L(\theta_L)V_L U_{L-1}(\theta_{L-1})...U_1(\theta_1)V_1$.
Here, $L$ is the number of parameters, $U_k(\theta_k)$ and $V_k$ are the $k$-th parametrized and fixed gates, respectively\textcolor{black}{, and $\theta_k$ is the $k$-th component of the parameter set $\vec{\theta}$}.  As an example of the cost function, in the celebrated variational quantum eigensolver (VQE)~\cite{peruzzo2014variational,mcclean2016theory}, one uses the expectation value of the Hamiltonian $H$, i.e., $\bra{\psi(\vec{\theta})} H \ket{\psi(\vec{\theta})}$. Typically, the parameters at $(j+1)$-th step $\vec{\theta}[j+1]$ is obtained by optimizing the cost function at the $j$-th step $C(\vec{\theta}[j])$ by using e.g., 
gradient descent methods. 
\textcolor{black}{The total number of iteration steps to update the parameters is defined as $M$.}
The other example of VQAs is variational quantum
simulation (VQS), which is used to simulate quantum dynamics such as Schr\"{o}dinger equation~\cite{li2017efficient,yuan2019theory}. By using the variational principles, it is possible to minimize the distance between the
\textcolor{black}{ideal state in the
exact evolution and the parametrized}
trial state, which provides us with the feasible update rule of parameters.  

\textcolor{black}{In variational algorithms, we should implement not only the original quantum circuit but also variant types of the original circuit.}
For example, in many variational algorithms, derivatives of quantum states, i.e., $\frac{\partial \ket{\psi(\vec{\theta})}}{\partial \theta_k}$ are used. They are generated from a different quantum circuit from the original ansatz circuit. 
\textcolor{black}{To discuss these cases in a general form, we denote the set of variational quantum circuits used in }
the algorithm as \textcolor{black}{$\{U^{(i)}_{\rm{AN}} (\vec{\theta})\}_{i=1}^G\equiv\{V^{(i)}_{L+1}U^{(i)}_{L}(\theta_L)V^{(i)}_{L}U^{(i)}_{L-1}(\theta_{L-1}) V^{(i)}_{L-1}\cdots  U^{(i)}_1(\theta_1)V^{(i)}_1\}_{i=1}^G$}, where $G$ is the number of
\textcolor{black}{variational quantum circuits including the original and variants.}
Accordingly, we denote the set of the observables measured in these quantum circuits as $\{\hat{A}^{(i)}_1,\hat{A}^{(i)}_2, \cdots, \hat{A}^{(i)}_{K^{(i)}}\}_{i=1}^G$, where 
\textcolor{black}{$\hat{A}^{(i)}$ is a Pauli matrix (or an operator made up of tensor products of the Pauli matrices) and}
$K^{(i)}$ is the number of observables measured in the $i$-th quantum circuits. \textcolor{black}{We will use these notation throughout this paper.}
We show \textcolor{black}{a prescription} about how
to implement the 
\textcolor{black}{conventional}
variational algorithms 
with these \textcolor{black}{notation}
in Appendix \ref{vqa}.

\section{Variational Secure Cloud  Quantum Computing}

We explain our protocol of the 
\textcolor{black}{variational secure cloud  quantum computing.}
Suppose that a client who has the ability to perform only single-qubit measurements hopes to access the 
NISQ computer of the server in a secure way. 
The main purpose of our scheme is to hide the information of the ansatz parameters $\vec{\theta}$ set by the client and output of the algorithm. In our scheme, 
the ansatz circuit to be implemented by the server is publicly announced beforehand.
Our scheme is 
efficient for the NISQ device that has a limited resource, because our scheme requires only \textcolor{black}{a single ancillary qubit independently of the number of qubits needed in the original NISQ algorithm.}
\textcolor{black}{These are in stark contrast with the original BQC.
In the
BQC, every information of the choice of the client is hidden~\cite{broadbent2009universal,PhysRevA.87.050301,PhysRevA.93.052307,Barz303,Greganti_2016}, while $3N$ qubits are approximately required to execute an algorithm using $N$ qubits.} 

\textcolor{black}{Throughout our paper,}
\textcolor{black}{
 we assume that the client has his/her own private space, and any information in the private space is not leaked to the outside.
This is the standard assumption in the quantum key distribution \cite{scarani2009security}}

\begin{figure}
  \includegraphics[width=0.5\textwidth]{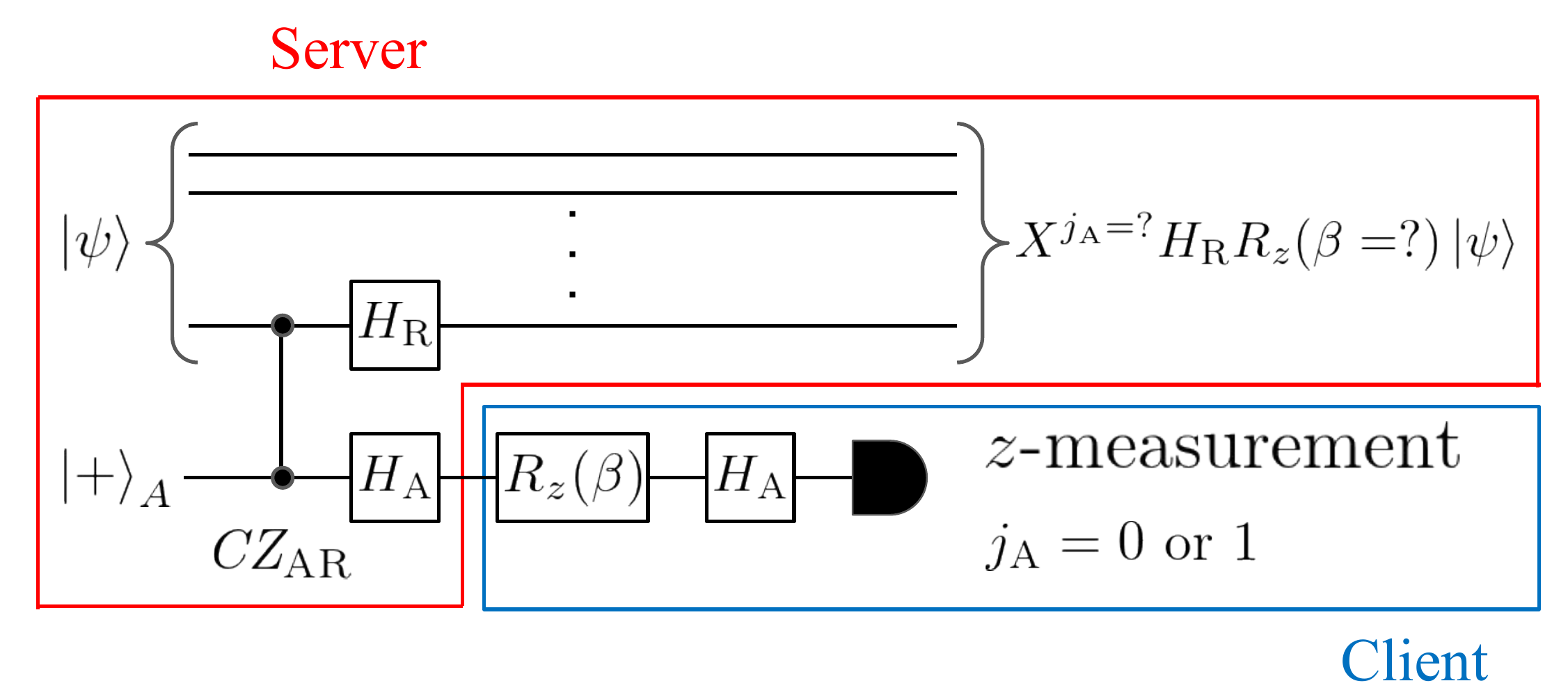}
  \caption{A quantum circuit to implement a single-qubit rotation by the client
  in our scheme.
  The circuit is the same as that in Fig.~\ref{fig:ADQC}.
  Firstly, the server entangles one register qubit with an ancillary qubit by the unitary operation  $E_{\mathrm{AR}}=H_{\mathrm{A}} H_{\mathrm{R}} \mbox{\it CZ}_{\mathrm{AR}}$.
  Secondly, the server sends the ancillary qubit to the client. \textcolor{black}{Thirdly}, the client performs a single-qubit rotation of $H_{\mathrm{A}} R_z(\beta)$ on the ancillary qubit, where $\beta$ is determined only by the client.
 \textcolor{black}{Finally, after the client measures the ancillary qubit in the $z$-basis,} $X^{j_{\mathrm{A}}} H_{\mathrm{R}} R_z(\beta)\ket{\psi}$ is generated for the register qubit.
  Since the client does not send any signals to the server, the server does not have any information about the rotation angle $\beta$ \textcolor{black}{and a measurement result $j_{\mathrm{A}}$, which is guaranteed by the no-signaling principle}.
  By repeating this process several times, arbitrary single-qubit rotations on a register qubit can be implemented.
  }
  \label{fig:single qubit rotation}
\end{figure}

The key 
of our protocol is to use the concept of the ADQC when the server runs the NISQ computing algorithm. We assume that the server has register qubits, and an ancillary qubit can be sent from the server to the client.
When the server needs to implement a single-qubit operation based on the ansatz, the server uses the single-qubit rotation scheme of the ADQC as shown in Fig. \ref{fig:single qubit rotation}. More specifically, the server performs a two-qubit gate $E_{\mathrm{AR}}$
between the register qubit (that we want to perform the single-qubit rotation) and the ancillary qubit, and sends the client the ancillary qubit to be measured by the client side.
The angle and axis of the single-qubit rotation are determined by the client. 
\textcolor{black}{With three sets of the rotation,  an arbitrary single-qubit rotation can be achieved in a register qubit (see Appendix~\ref{detailed ADQC}).}

Moreover, by performing a single-qubit rotation on every register qubit in this way, we have byproduct operators of 
$X^{j_1+j_3}Z^{j_2}$ on every register qubit
as shown in Eq. (\ref{arbitraty single gate}).
\textcolor{black}{ 
It is known that, when Pauli matrices or an identity operator are randomly implemented on a quantum state (see Sec. 8.3.4 in Ref. \cite{nielsen2002quantum}), the state becomes completely mixed. This means that the byproduct operators make the state completely mixed for the server.}
Due to this property, any measurements on the server side provide random outcomes 
\textcolor{black}{if the server side does not have any information of the client's dataset}, which is helpful for the client to hide the output of the algorithm.
\textcolor{black}{The security of our scheme can be also interpreted as follows.}
\textcolor{black}{
During the implementations of these gates in our scheme,
 the gate operations executed by the server do not depend on the ansatz parameters. 
\textcolor{black}{Moreover,}
the client does not send any information to the server \textcolor{black}{during our protocol}. Therefore, the server cannot find the parameters of the ansatz circuit set by the client.
Such security is
guaranteed by the no-signaling principle  \cite{popescu1994quantum,morimae2013blind}.}

\begin{figure}
  \begin{minipage}[]{1\linewidth}
  \centering
  \subcaption{\raggedright}
  \includegraphics[clip,width=1\textwidth]{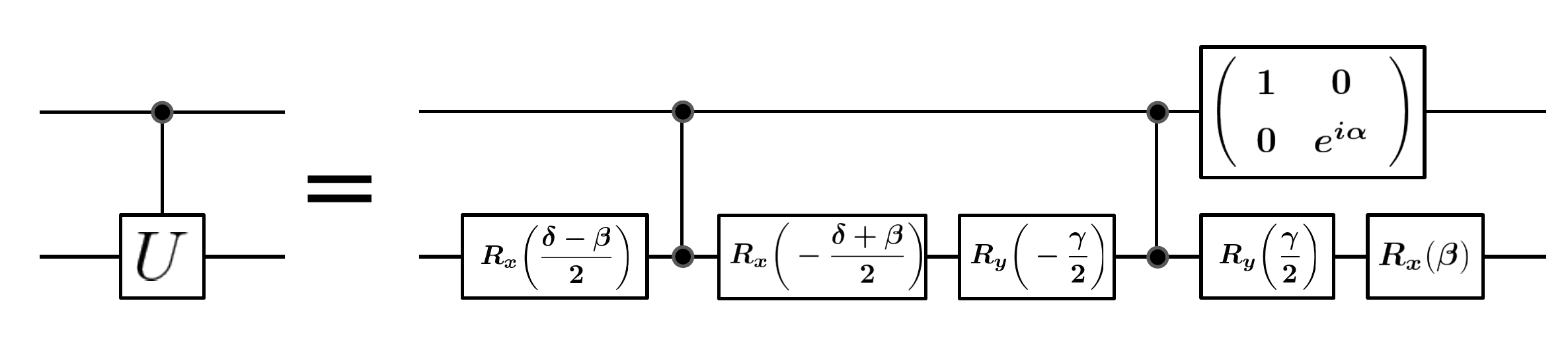}
  \label{fig:two qubit rotation}
  \end{minipage}\\
  \begin{minipage}[]{1\linewidth}
  \centering
  \subcaption{\raggedright}
  \includegraphics[clip,width=1\textwidth]{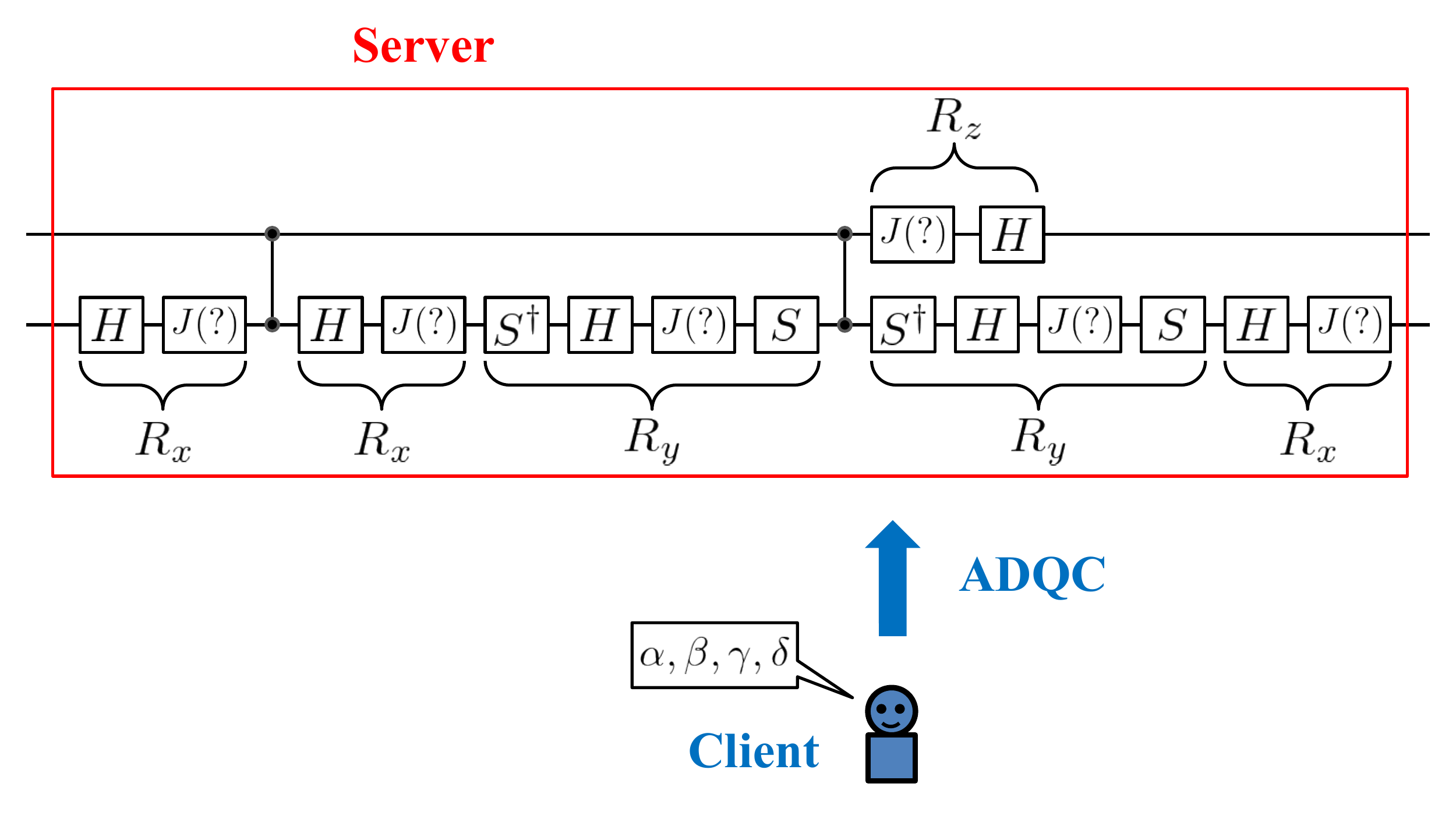}
  \label{ourtwoqubitgate}
  \end{minipage}
  \caption{
  \textcolor{black}{An implementation of two-qubit gates in our scheme.
  (a) An equivalent circuit with a controlled-gate operation. 
  We can \textcolor{black}{decompose} an arbitrary two-qubit gate into several 
  \textcolor{black}{gates such as single-qubit rotations (including parameters) and controlled-$Z$ gates, where}
  we need to 
  \textcolor{black}{choose appropriate parameters of} $\alpha, \beta, \gamma$, and $\delta$ for the equivalence. 
  }
  (b) A quantum circuit to implement an arbitrary controlled-gate operation by the client while the rotation parameters are hidden to the server
  in our scheme. The basic structure of the
  circuit is the same as that in (a),
  \textcolor{black}{where $S$ denotes a phase gate}.
  \textcolor{black}{The Hadamard, the controlled-$Z$, and the phase gates are implemented by the server in the register qubits.}
  An important point is that every single-qubit rotation in the circuit should be performed by the client 
  in the same way as described in Fig.~\ref{fig:single qubit rotation}. 
  In this case, no-signaling principle guarantees that the rotation parameters ($\alpha$, $\beta$, $\gamma$, and $\delta$)
  cannot be inferred by any operation on the server.
  }
\end{figure}

When the server needs to perform a two-qubit gate based on the ansatz with a specific angle, we adopt a quantum circuit shown in Fig. \ref{fig:two qubit rotation}. 
\textcolor{black}{The point is that an arbitrary two-qubit gate can be decomposed by arbitrary single-qubit gates and controlled-$Z$ gates.}
We combine the single-qubit rotations in the ADQC \textcolor{black}{with} two controlled-$Z$ gates as shown in Fig. \ref{ourtwoqubitgate}. In this case, the angles of the two-qubit gates can be determined by the client
\textcolor{black}{because the angle of the single-qubit gate can be specified just by the client. Similar to the case of the single-qubit gates, the no-signaling principle guarantees that the server does not obtain any information about the ansatz parameters during the implementation of the two-qubit gates.}

\begin{figure*}
  \includegraphics[width=0.8\textwidth]{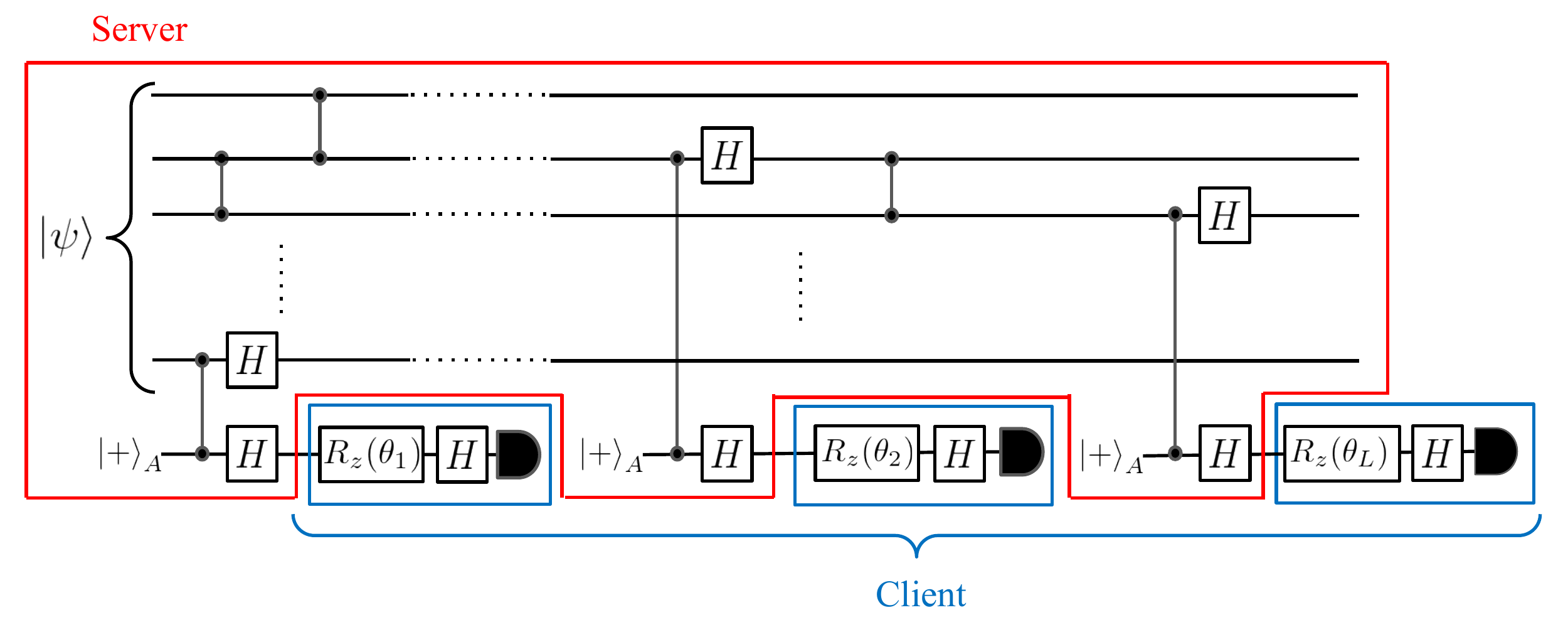}
  \caption{A quantum circuit to implement \textcolor{black}{our variational secure cloud quantum computing.}
  The NISQ algorithm requires the parameters $\{\theta _j\}_{j=1}^L$ to change the ansatz circuit in a variational way.
  The server implements gate operations that do not depend on the parameters, and sends the ancillary qubit to the client.
  On the other hand,
  the client can specify the parameters by changing the measurement angles on the ancillary qubits sent from the server. Importantly, in our scheme, the client does not send any signal to the server, and thus the server does not know the parameters set by the client\textcolor{black}{, due to the no-signaling principle}.
  }
  \label{ourconceptone}
\end{figure*}

\textcolor{black}{The combinations of the single-qubit gates and two-qubit gates in our architecture are shown in 
Fig. \ref{ourconceptone}.
The server performs only Hadamard gates, phase gates, and controlled-$Z$ gates, \textcolor{black}{which are clifford gates.}
Therefore, 
\textcolor{black}{when the server measures the observables in the register qubits and sends the measurement results to the client,}
the client can effectively remove the effect of the byproduct operators by changing the interpretation of the measurement results (see Appendix~\ref{detailed ADQC}).}


\begin{figure}
  \includegraphics[width=0.5\textwidth]{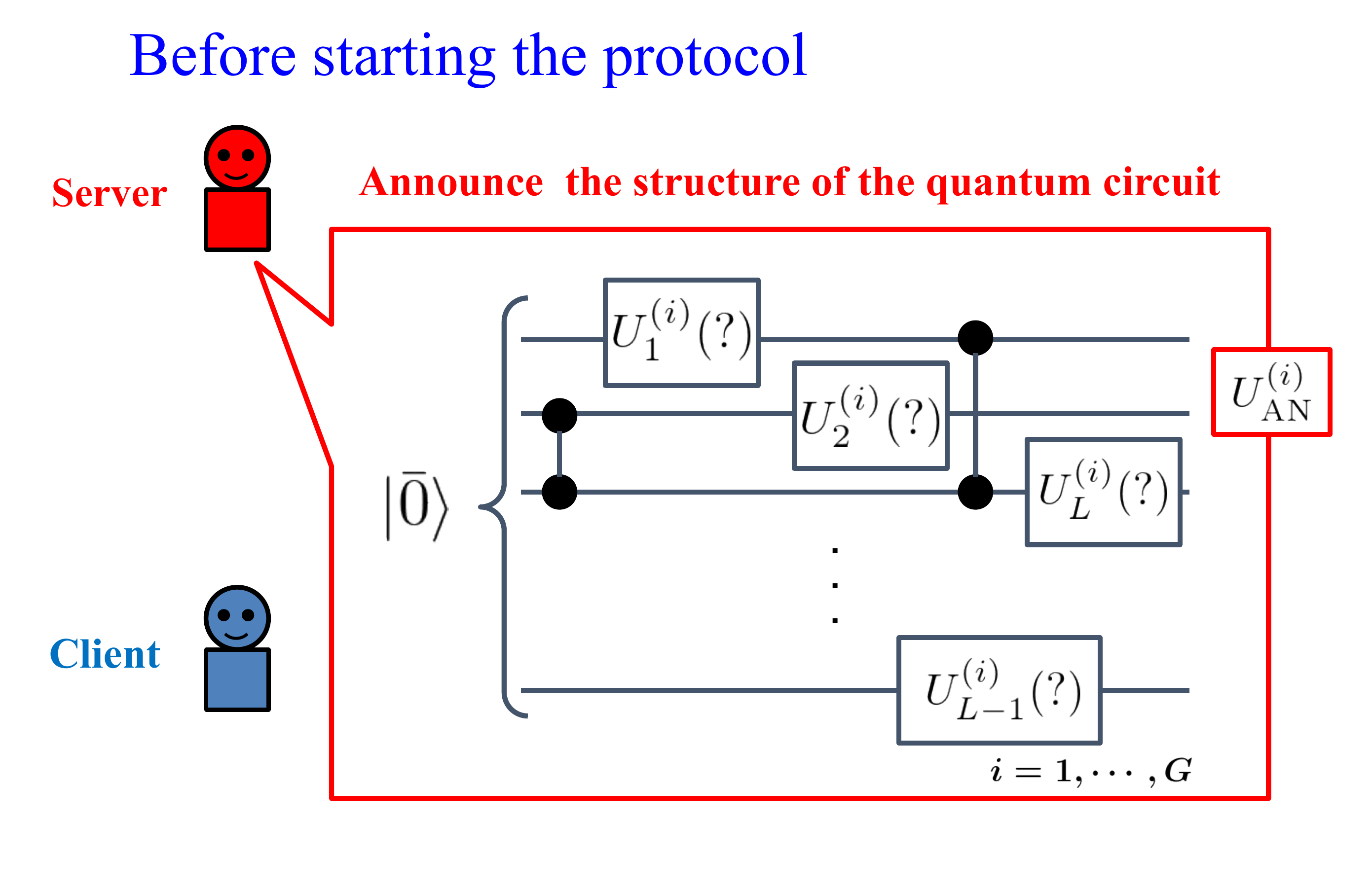}
  \caption{Before the client starts the protocol, the server broadcasts the information about their quantum circuit. This includes the set of unitary operations $\{U^{(i)}_{\rm{AN}}\}_{i=1}^G$, the set of the observables  $\{\hat{A}^{(i)}_1,\hat{A}^{(i)}_2, \cdots, \hat{A}^{(i)}_{K^{(i)}}\}_{i=1}^G$ to be measured, the repetition numbers $\{N^{(i)} \}_{i=1}^G$ for the quantum circuits, initial states $\{\ket{\psi^{(i)}(\vec{\theta}[0])}\}_{i=1}^G$, and the total number $M$ of iteration steps to update the parameters. The client implements the NISQ algorithm based on this information.
  }
    \label{ourconcepttwo}
\end{figure}

\textcolor{black}{Before the client performs the secure cloud NISQ computation,}
\textcolor{black}{the server publicly announces} the set of unitary operators $\{U^{(i)}_{\rm{AN}}\}_{i=1}^G$, the set of the observables  $\{\hat{A}^{(i)}_1,\hat{A}^{(i)}_2, \cdots, \hat{A}^{(i)}_{K^{(i)}}\}_{i=1}^G$, the repetition numbers $\{N^{(i)} \}_{i=1}^G$
for sampling with the quantum circuits, initial states $\{\ket{\psi^{(i)}(\vec{\theta}[0])}\}_{i=1}^G$, 
\textcolor{black}{$L$ (the number of \textcolor{black}{variational} parameters), 
$M$ (the total number of iteration steps \textcolor{black}{for VQAs}), and $G$ (the number of variants of variational quantum circuits),}
 as shown in Fig. \ref{ourconcepttwo}.
 
\begin{figure*}
  \includegraphics[width=1\textwidth]{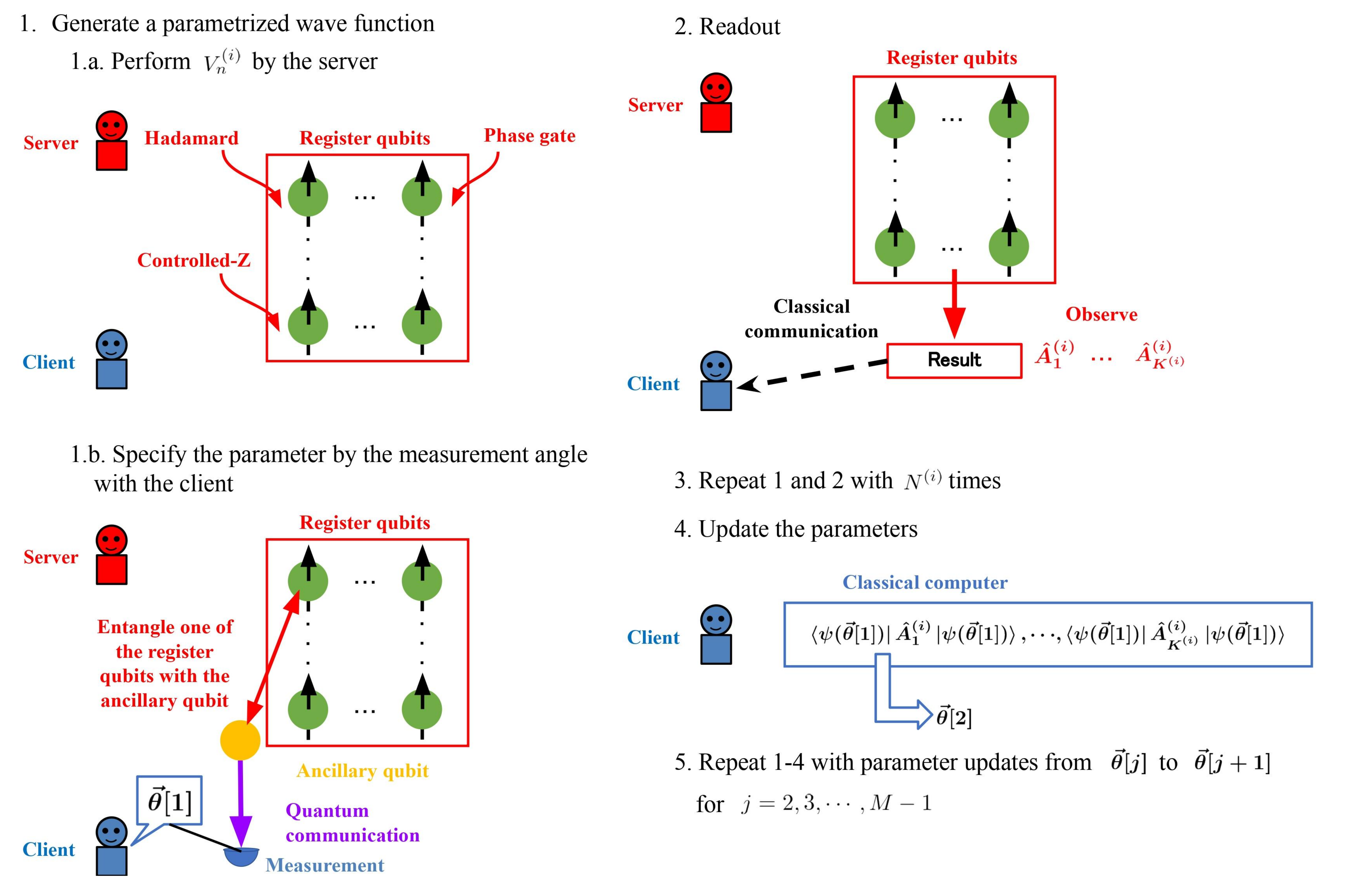}
  \caption{
  The sequence of our scheme to implement a NISQ algorithm with a parameter set of 
  \textcolor{black}{$\vec{\theta}$}
  in 
  \textcolor{black}{a variational secure cloud quantum computing}.
  (1) The server sequentially performs 
  the unitary operations $\{U^{(i)}_{\rm{AN}}\}_{i=1}^G$ for the register qubits, \textcolor{black}{where $G$ denotes the number of the quantum circuits to be performed.} \textcolor{black}{
  (1.a) The server implements a unitary (non-parameterized) operation \textcolor{black}{$V^{(i)}_n$ for $n=1,2,\cdots, L+1$}; a Hadamard or a controlled-$Z$, \textcolor{black}{or a phase gate,} on the register qubits. 
  (1.b) }
  The server entangles a register qubit with an ancillary qubit and sends the ancillary qubit to the client in the same way as Fig.~\ref{fig:single qubit rotation}. The client measures the ancillary qubit sent from the server, where the client specifies a measurement angle based on the initial parameters
  $\vec{\theta}[1]$.
(2) The server 
measures
$\hat{A}^{(i)}_1,\hat{A}^{(i)}_2, \cdots,$ and $\hat{A}^{(i)}_{K^{(i)}}$ for $i=1,2,\cdots, G$
and sends the results to the client by the classical communication. 
(3) For each $\{U^{(i)}_{\rm{AN}}\}_{i=1}^G$,
the server and the client
repeat these two steps  
$\{N^{(i)}\}_{i=1}^G$ times, and then the client obtains expectation values of 
$\{\hat{A}^{(i)}_1,\hat{A}^{(i)}_2, \cdots, \hat{A}^{(i)}_{K^{(i)}}\}_{i=1}^G$ with 
\textcolor{black}{$\{\ket{\psi^{(i)}(\vec{\theta}[1])}\}_{i=1}^G$}. 
(4) The client updates the parameters as 
\textcolor{black}{$\vec{\theta}[2]$}
by processing the measurement data with classical computation. 
(5)
The server and the client
repeat these four steps with $M-2$ times updating the parameters from 
\textcolor{black}{$\vec{\theta}[j]$ to $\vec{\theta}[j+1]$}
for $j=2,3,\cdots, M-1$, and the client obtains the output. Since the client does not send any signals to the server during the computation, the server cannot obtain any information about 
\textcolor{black}{$\vec{\theta}[1]$, $\vec{\theta}[2]$, $\dots$, $\vec{\theta}[M]$\textcolor{black}{, because of the no-signaling principle}}.
}
    \label{ourconceptthree}
\end{figure*}

\textcolor{black}{We summarize our scheme in Fig.~\ref{ourconceptthree} as follows.}
\begin{enumerate}
\item 
Adopting the quantum circuits of $\{U^{(i)}_{\rm{AN}}\}_{i=1}^G$, the server and client implement these unitary operations
to generate the trial wave functions 
\textcolor{black}{$\{\ket{\psi^{(i)}(\vec{\theta}[1]
)}\}_{i=1}^G$}. 
Here, parametrized single- and two-qubit gates should be implemented in the specific ways as described in Figs.~\ref{fig:single qubit rotation} and \ref{ourtwoqubitgate}, respectively. 
\textcolor{black}{More specifically, the server performs operations,
\textcolor{black}{ such as the Hadamard, the controlled-$Z$, and the phase gates}
\textcolor{black}{(1.a in Fig.~\ref{ourconceptthree})}, while the client specifies the \textcolor{black}{measurement angles} 
\textcolor{black}{(1.b of Fig.~\ref{ourconceptthree}).}}
\textcolor{black}{
We do not need to prepare 
$\{\ket{\psi^{(i)}(\vec{\theta}[1])}\}_{i=1}^G$
simultaneously 
\textcolor{black}{by using $G$ quantum computers,}
but we can prepare and measure these in sequence \textcolor{black}{by using a single quantum computer}, \textcolor{black}{similar to the standard VQA for NISQ devices (see Appendix \ref{vqa})}.}

\item 
The server measures the states of the register qubits with  $\{\hat{A}^{(i)}_1,\hat{A}^{(i)}_2, \cdots, \hat{A}^{(i)}_{K^{(i)}}\}_{i=1}^G$, and sends the results to the client with classical communications.

\item For the sampling, the server and client repeat
the first \textcolor{black}{and the second}
steps with
$\{N^{(i)}\}_{i=1}^G$ times for each state
\textcolor{black}{
$\{\ket{\psi^{(i)}(\vec{\theta}[1])}\}_{i=1}^G$
}
so that the client should obtain the expectation values of  
$\{\hat{A}^{(i)}_1,\hat{A}^{(i)}_2, \cdots, \hat{A}^{(i)}_{K^{(i)}}\}_{i=1}^G$.
\textcolor{black}{
When the observables are measured, the effect of the byproduct operators can be canceled out by the client (see Appendix~\ref{detailed ADQC}).}

\item By processing the measurement results with a classical computer at the client side, the client updates the parameters and obtains
\textcolor{black}{$\vec{\theta}[2]=(\theta_1[2], \cdots, \theta_L[2])^T$} 
for the next step. 

\item The client and the server repeat
the 
\textcolor{black}{steps 1-4}
\textcolor{black}{$(M-2)$} 
times
with $\{U^{(i)}_{\rm{AN}}\}_{i=1}^G$ 
and 
\textcolor{black}{$\vec{\theta}[j]$,}
where 
classical computation based on the results at the $j$-th step
provides the client with
the updated parameters of 
\textcolor{black}{$\vec{\theta}[j+1]$}
for $j=2,3, \cdots ,M-1$. 
\textcolor{black}{The client finally obtains desired results in a secure way from the server.}
\end{enumerate}

\textcolor{black}{
As a physical implementation, the register qubits can be the solid-state systems that interact with a photon, and the ancillary qubit can be an optical photon that transmits to a distant place. 
\textcolor{black}{We implicitly assumed
that the photon loss would be negligible during the transmission in the discussion above.
}
}

Finally, we discuss the effect of photon loss 
on our scheme.
When the server sends the client an ancillary qubit that corresponds to an optical photon, there is a possibility that the photon can be lost during the transmission.
\textcolor{black}{In principle, if the server and the client have quantum memories, they can share a Bell pair under the effect of photon loss
by repeating the entanglement generation process until success~\cite{benjamin2006brokered}, and they can use the Bell pairs to perform our gate operations in a deterministic way. In this case, the client needs to ask the server to send the photons again and again, depending on how many times the photon is lost~\cite{benjamin2006brokered}.}
However, in order to apply the no-signaling principle, \textcolor{black}{the client is not allowed to send the server any information. This means that} the client cannot ask the server to send the photon again. 
\textcolor{black}{So we cannot adopt the repeat-until-success strategy with quantum memories.}

Thus, we assume that the client adopts the \textcolor{black}{observation}
results of the readout of the register qubits by the server
\textcolor{black}{only when all photons are successfully transmitted to the client during}
the computation. 
In this case, the probability of the no photon loss during the computation exponentially decreases as the number of sending photons increases. The number of required photons 
sent to
the client
can be determined by the number of the tunable parameters used in the ansatz circuit. 
When $U^{(i)}_{\mathrm{AN}}$ is composed of $n^{(i)}_{\rm{single}}$ single-qubit operations and $n^{(i)}_{\rm{two}}=L-n^{(i)}_{\rm{single}}$ two-qubit operations, the necessary number $N_{\rm{ph}}^{(i)}$ of the photons to send the client is at most 
$N^{(i)}_{\rm{ph}}=3n^{(i)}_{\rm{single}} + 6n^{(i)}_{\rm{two}}$ \textcolor{black}{as shown in Eq.~\ref{arbitraty single gate} and Figs.~\ref{fig:single qubit rotation} and \ref{ourtwoqubitgate}}. 
The probability for all the photons to be detected by the client is $(1-p_{\rm{loss}})^{ N_{\rm{ph}}^{(i)}}$\textcolor{black}{, where $p_{\rm{loss}}$ is a photon loss probability for a single transmission}. 
Therefore, the repetition number
\textcolor{black}{$N^{(i)}$} with the photon loss
should be \textcolor{black}{set to be}
much larger than \textcolor{black}{$N_{\rm{ideal}}^{(i)}/(1-p_{\rm{loss}})^{N_{\rm{ph}}^{(i)}}$}, where \textcolor{black}{$N_{\rm{ideal}}^{(i)}$} denotes the required number of repetition with no photon loss. To keep \textcolor{black}{$N^{(i)}$} within a reasonable amount, $p_{\rm{loss}}$ should be smaller than $1\%$ under the assumption that $N_{\rm{ph}}^{(i)}$ is around a few hundreds.

We could overcome such a problem due to the recent experimental and theoretical developments of quantum repeating technology.
The best single-photon detector in optics has $99\%$ efficiency~\cite{lita2010superconducting,fukuda2011titanium,kuzanyan2018three}. Microwave quantum repeater with a short distance such as 100 m has been proposed~\cite{xiang2017intracity}, and 
a qubit can catch a microwave photon with $99.4\%$ absorption efficiency
in the microwave regime~\cite{wenner2014catching}. 
Also, there are proposals 
to physically move the solid-state qubit~\cite{schaffry2011proposed,devitt2016high} for distributed quantum computation or a quantum repeater
\textcolor{black}{Through the combination of these protocols and a long-lived quantum memory such as a nuclear spin~\cite{saeedi2013room,aslam2017nanoscale}, the ancillary solid-state qubits might be carried to the client without the problems of the photon loss.}

\section{Conclusion}
In conclusion, we proposed a noisy intermediate-scale quantum (NISQ) computing with security inbuilt.
\textcolor{black}{The main targets of our scheme are variational quantum algorithms (VQAs), which involve parameters of an ansatz to be optimized by minimizing a cost function.}
We considered a circumstance that a client with a limited ability to perform quantum operations hopes to access a NISQ device possessed by a server and the client tries to avoid leakage of the information about the quantum algorithm that he/she runs.
Importantly, \textcolor{black}{the naive application of the previously known} blind quantum computation (BQC)~\cite{morimae2013blind}
 requires around $3N$ qubits~\cite{raussendorf2001one,raussendorf2003measurement,walther2005experimental}, where $N$ denotes the number of the qubits to run the quantum algorithm in the original architecture. \textcolor{black}{That} may not be suitable for the NISQ devices with the limited number of qubits. 
\textcolor{black}{Our proposal}
is more efficient in the sense that we use 
\textcolor{black}{a single}
\textcolor{black}{ancillary qubit and $N$ register qubits required in the original NISQ algorithm.}
\textcolor{black}{In VQAs, we use a parametrized trial wave function,} and our scheme prevents the information about the parameters from the leakage to the server. We rely on the no-signaling principle to guarantee security. Our scheme paves the way for new applications of the NISQ devices. 

Y. S. and Y. T.  contributed to this
work equally.
This work was supported by Leading Initiative for
Excellent Young Researchers MEXT Japan and JST
presto (Grant No. JPMJPR1919) Japan.  This work was supported by MEXT Quantum
Leap Flagship Program (MEXT Q-LEAP) (Grant No.
JPMXS0120319794, JPMXS0118068682), JST ERATO (Grant No. JPMJER1601),
JST [Moonshot R\&D--MILLENNIA Program] Grant No. JPMJMS2061, MEXT Quantum Leap Flagship Program (MEXT Q-LEAP) Grant No. JPMXS0118067394, and \textcolor{black}{S.W. was supported by Nanotech CUPAL,National Institute of Advanced Industrial Science and Technology (AIST).
This paper was partly based on results obtained from a project, JPNP16007, commissioned by the New Energy and Industrial Technology Development Organization (NEDO), Japan. }

\appendix{}

\textcolor{black}{
\section{\textcolor{black}{Detailed ancilla-driven quantum computation}}\label{detailed ADQC}
\subsection{Arbitrary single-qubit rotation}
We describe a way to implement an arbitrary single-qubit rotation.
Any single-qubit rotation $U$ can be represented by $U=R_z(\beta')R_x(\gamma')R_z(\delta')$, where $R_x$ denotes a rotation about the $x$-axis, and
$\beta'$, $\gamma'$, and $\delta'$ denote the rotation angles about the corresponding axis.  
 Defining $J(\beta)\equiv HR_z(\beta)$, one can rewrite $U$ as $U=J(\beta)J(\gamma)J(\delta)$, where we choose $\beta$, $\gamma$, and $\delta$ to satisfy $R_z(\beta)R_x(\gamma)R_z(\delta)=HU$. As we explained, one can implement the single-qubit rotation of $X^j H_{\mathrm{R}} R_z(\beta)\ket{\psi}$ on the register qubit by the coupling with an ancillary qubit and a subsequent measurement. Therefore, three sequential operations of this type of the single-qubit rotation provide us with the following operation
\footnotesize
\begin{eqnarray}
&&\Big{(}X^{j_3}H_{\mathrm{R}} R_z((-1)^{j_2}\beta)\Big{)}\Big{(}X^{j_2}H_{\mathrm{R}} R_z((-1)^{j_1}\gamma)\Big{)}\Big{(}X^{j_1}H_{\mathrm{R}} R_z(\delta)\Big{)}
\nonumber \\
    &=&X^{j_3}J((-1)^{j_2}\beta)X^{j_2}J((-1)^{j_1}\gamma)X^{j_1}J(\delta) \nonumber \\ 
    &=&(-1)^{j_1 \cdot j_2}X^{j_1+j_3}Z^{j_2}J(\beta)J(\gamma)J(\delta)
\label{arbitraty single gate}
\end{eqnarray}
\normalsize
where $j_i$ denotes the result of the $i$-th measurement on the ancillary qubits. For the implementation of this operation, we change the rotation angle of the ancillary qubit depending on the previous measurement results. 
Equation.~(\ref{arbitraty single gate}) involves the byproduct operator $X^{j_1+j_3}Z^{j_2}$. However, as long as we measure the qubit in a computational basis for the readout, the byproduct operators just flip the measurement result from $0$ to $1$ or vice versa, and so we can effectively remove the byproduct operators from the states
\textcolor{black}{by changing the interpretation of the measurement results.}
\subsection{Two-qubit gate between the register qubits}
We explain a way to perform the controlled-$Z$ gate on the two register qubits $R$ and $R'$ in the ADQC. Firstly, we implement $E_{\mathrm{AR}}$ on the ancillary qubit (prepared in the state $\ket{+}_{\mathrm{A}}$) and the register qubit $R$, and subsequently perform $E_{\mathrm{AR'}}$ on the ancillary qubit and the other register qubit $R'$.
Secondly, one measures the ancillary qubit in the y-basis. 
\textcolor{black}{These operations are equivalent to the controlled-$Z$ gate, up to local operations.}
}
\textcolor{black}{
When we perform several single-qubit gates and two-qubit gates, the byproduct operators are applied as $U_{\Sigma} U_{\rm{ideal}}|\overline{0}\rangle$, where $U_{\Sigma}$ denotes the total byproduct operators and $U_{\rm{ideal}}$ denotes the unitary operations that we aim to implement. Again, when one measures observables of Pauli matrices (or a tensor product of Pauli matrices), one can effectively remove the byproduct operators from the states
\textcolor{black}{by changing the interpretation of the measurement results.}
}

    \section{VQA for NISQ devices}\label{vqa}
We show \textcolor{black}{a prescription} about how to implement the
\textcolor{black}{conventional}
variational algorithms 
with  \textcolor{black}{our notation. We}
\textcolor{black}{prepare a parametrized wave function on a quantum circuit $\ket{\psi(\vec{\theta})}$ with the variational parameters $\vec{\theta}$ to be optimized by minimizing a cost function $C(\vec{\theta})$ tailored to a problem.}
Firstly, with the quantum circuits of $\{U^{(i)}_{\rm{AN}}\}_{i=1}^G$, we realize parametrized wave functions of $N$-qubits 
\textcolor{black}{$\{\ket{\psi^{(i)}(\vec{\theta}[1])}\}_{i=1}^G$,} 
where 
\textcolor{black}{$\ket{\psi^{(i)}(\vec{\theta}[1])}\equiv V^{(i)}_{L+1}U^{(i)}_{L}(\theta_{L}[1])V^{(i)}_{L} \cdots U^{(i)}_{1}(\theta_1[1])V^{(i)}_{1}
\ket{\bar{0}}$} \textcolor{black}{with  $\ket{\bar{0}} \equiv \bigotimes_{i=1}^{N}\ket{0}$}
denotes the wave function, 
\textcolor{black}{$\vec{\theta}[1]=(\theta_1[1], \cdots, \theta_L[1])^T$}
is a vector of the parameters and $\{\ket{\psi^{(i)}(\vec{\theta}[0]
)}\}_{i=1}^G$ are initial states, and we measure the state of the wave function with observables of  $\{\hat{A}^{(i)}_1$, $\hat{A}^{(i)}_2$,
$\cdots$, $\hat{A}^{(i)}_{K^{(i)}}\}_{i=1}^G$. 

Secondly, for the sampling, we repeat the first step to obtain expectation values of
$\{\hat{A}^{(i)}_1,\hat{A}^{(i)}_2, \cdots, \hat{A}^{(i)}_{K^{(i)}}\}_{i=1}^G$ with 
\textcolor{black}{$\{\ket{\psi^{(i)}(\vec{\theta}[1])}\}_{i=1}^G$}.
Thirdly, based on the expectation values, we implement
a classical algorithm so that we can obtain updated
parameters 
\textcolor{black}{$\vec{\theta}[2]$}
for the next quantum circuits, where we typically use a gradient method to make the cost function smaller.
\textcolor{black}{For example, we use  $\vec{\theta}[j+1]=\vec{\theta}[j]-\alpha {\rm{grad}} C(\vec{\theta}[j])$ for the gradient method.}

Finally, we 
repeat
the first, second,  and third steps $M-2$ 
times
with $\{U^{(i)}_{\rm{AN}}\}_{i=1}^G$ 
and 
\textcolor{black}{$\vec{\theta}[k]$,}
where 
classical computation based on the results at the $k$-th step
provides the updated parameters of 
\textcolor{black}{$\vec{\theta}[k+1]$}
for $k=2,3, \cdots ,M-1$. 
These processes provide us with an output of the algorithm.

\bibliographystyle{apsrev4-1}
\bibliography{blindnisq}
\end{document}